\documentclass[rnote]{aa} 

\usepackage{graphicx}
\usepackage{txfonts}
\begin{document}
   \title{Current data on the globular cluster Palomar 14 are not inconsistent with MOND}

   \author{G. Gentile
          \inst{1,2}
          \and
          B. Famaey
          \inst{3,4}
          \and
          G. Angus\inst{5,6}
          \and
          P. Kroupa
          \inst{4}
          }

   \institute{Institut d'Astronomie et d'Astrophysique, Universit\'e Libre de Bruxelles, CP 226, Bvd du Triomphe, B-1050, Bruxelles, Belgium
         \and
             Sterrenkundig Observatorium, Universiteit Gent, Krijgslaan 281, B-9000 Gent, Belgium\\
              \email{gianfranco.gentile@ugent.be}
         \and
         Observatoire Astronomique,  Universit\'e de Strasbourg, CNRS UMR 7550, F-67000 Strasbourg, France
\\ \email{benoit.famaey@astro.unistra.fr}
         \and
         AIfA, Universit\"at Bonn, D-53121 Bonn, Germany 
         \\ \email{pavel@astro.uni-bonn.de}
         \and
          Dipartimento di Fisica Generale ``Amedeo Avogadro", Universit\`a degli studi di Torino, Via P. Giuria 1, I-10125, Torino, Italy \\ \email{angus@ph.unito.it}
         \and
          Istituto Nazionale di Fisica Nucleare (INFN), Sezione di Torino, Torino, Italy}

   \date{Received September 15, 1996; accepted March 16, 1997}

 
  \abstract
   {Certain types of globular clusters have the very important property that the 
predictions for their kinematics in the Newtonian and modified Newtonian dynamics (MOND) contexts are 
divergent.}
   {Here, we caution the recent claim
that the stellar kinematics data (using 17 stars) of the globular cluster Palomar 14 are inconsistent with MOND.}
   {We compare the observations to the theoretical predictions using a Kolmogorov-Smirnov
test, which is appropriate for small samples. }
   {We find that, with the currently available data, the MOND prediction for the velocity distribution can only be
excluded with a very low confidence level, clearly
insufficient to claim that MOND is falsified. }
  {}


   \maketitle
%

\section{Introduction}

A plethora of observational data on various astronomical scales
seem to support the idea that the amount of visible matter in the Universe
is several times smaller than the total amount of matter (e.g. Hinshaw et 
al. 2009). The current
paradigm of structure formation and evolution in the Universe is known as the 
$\Lambda$ Cold Dark Matter ($\Lambda$CDM) model. 
However, as long as the dark matter 
particle has not been discovered (and its cosmological abundance confirmed), it is
worth considering alternative theories to explain the current data.

For instance, MOND (modified Newtonian dynamics) was proposed by Milgrom 
(1983) as an alternative to galactic dark matter. MOND stipulates that
below a certain gravitational acceleration $a_0 \sim 1.2 \times 
10^{-8}$ cm s$^{-2}$ the actual gravitational acceleration $g$ is 
stronger than expected in Newtonian gravity ($g_{\rm N}$). 
Asymptotically, in MOND it reaches the value $\sqrt{g_{\rm N} a_0}$
for $g_{\rm N} \ll a_0$. This allows it to naturally explain various galaxy scaling relations (e.g., Faber \& Jackson~1976, Tully \& Fisher~1977, McGaugh et al.~2000, McGaugh~2004, Gentile~2008, Donato et al.~2009, Gentile et al.~2009).

The effects of MOND and dark matter are however often rather degenerate and 
model-dependent since the gravitational potential predicted by MOND can almost always be attributed  by a Newtonist to an {\it ad hoc} dark matter distribution. Objects for which the predictions of the two theories
are unambiguously different are unfortunately rare. We note that galaxy clusters are not good discriminant tests: indeed, in galaxy clusters, the acceleration predicted by MOND is not large enough (e.g., Sanders~1999), but some form of hot dark matter can be added within the MOND context to make the 
data consistent with the predictions (e.g., Angus et al.~2009). If there were systems where MOND predicted more gravity than observed, they would make a strong case against MOND.

One example of such objects that should be (almost) devoid of cold dark matter in the $\Lambda$CDM cosmological model, and for which MOND predicts 
significantly stronger gravity than that attributable to visible matter,
are tidal dwarf galaxies (TDGs). However, from the
observations of three young TDGs around NGC~5291, Gentile et al. (2007) found that MOND remarkably fits the data with zero free parameters, whereas CDM fails to explain them. 

Apart from TDGs, another type of object that has recently been
put forward as a discriminant test for CDM and MOND are
the globular clusters of our own Galaxy (Baumgardt et al.~2005). Of particular interest are 
those that are diffuse and distant from the Milky Way, to ensure that the 
internal acceleration probes the deep MOND regime (which is not always 
necessarily the case: see NGC~2419 in Baumgardt et al. 2009) and that, 
simultaneously, the gravitational acceleration due to the Milky Way (external
field) is weak enough. In MOND the deviation from a Newtonian behaviour 
should start appearing around $a_0$, whereas in the ``Newtonian
gravity plus CDM'' picture no discrepancy is expected since globular clusters 
should contain (almost) no cold dark matter.

In a recent paper, Jordi et al.~(2009) analysed the data of 17 stars
from the globular cluster Palomar~14, and claimed that (within the 
assumption of Palomar~14 being on a circular orbit) MOND is inconsistent
with the observed velocity dispersion: the MOND prediction is too high,
whereas the data are consistent with the Newtonian prediction (with no
dark matter).
They discuss two separate cases, depending on the inclusion or not of
a star (Star 15), which based on its line-of-sight velocity is not a definite
member of the cluster. However, given such a small sample size, an appropriate test should be used to discern between the hypotheses. An example of such
a test is the Kolmogorov-Smirnov (KS) test (e.g. Soong 2004). An illustrative example of the uncertain value of the velocity dispersion measured with a small number of stars is NGC~2419: the original analysis by Olszewski et al.~(1993) yielded a velocity dispersion for 12 stars of 2.7$\pm$0.8~km$\, {\rm s}^{-1}$, which was re-evaluated by Baumgardt et al.~(2009) with 40 stars to be 4.1$\pm$0.5~km$\, {\rm s}^{-1}$, i.e. about two sigma greater than the original. This gives a perfectly reasonable M/L in MOND, although the velocity dispersion profile, if confirmed, could be a problem for MOND (Sollima \& Nipoti 2009).
This will allow a very interesting test of MOND when data points at larger radii will be obtained in NGC~2419.
In any case, this is an excellent example of an underestimation of the true velocity dispersion due to an originally too small sample size.

Therefore, the question we ask in the present research note is the following:
do the Palomar~14 data really exclude MOND, given the small number of 
stars used in the analysis of Jordi et al.~(2009)? We then also investigate the minimum number of stars needed to exclude MOND in a globular cluster similar to Palomar~14. Or, equivalently, we ask how many globular clusters like Palomar~14 would be needed to exclude MOND.

\begin{table}   
\centering   
\begin{scriptsize}   
\centering   
\caption[]{Velocities and cumulative distribution functions (observed
and Gaussian, with and without Star 15). The star names are taken from
Hilker (2006) when present, otherwise from Harris \& van den Bergh (1984) 
and Holland \& Harris (1992).
}   
\vspace{0.3cm}   
\label{tab-cdf}   
\begin{tabular} {l l l l l l}    
\hline     
\hline     
Name     &velocity       & obs cdf      & Gauss cdf    & obs cdf          & Gauss cdf      \\   
         & (km s$^{-1}$) & with Star 15 & with Star 15 & w/o Star 15      & w/o Star 15    \\   
\hline     
15       &  69.99        & 0.059       & 0.034       & -                &  -                 \\   
8        &  71.38        & 0.118       & 0.234       & 0.063            &  0.209             \\   
3        &  71.75        & 0.177       & 0.332       & 0.125            &  0.302             \\   
14       &  71.80        & 0.235       & 0.347       & 0.188            &  0.316             \\   
12       &  71.83        & 0.294       & 0.356       & 0.250            &  0.324             \\   
HH042    &  71.94        & 0.353       & 0.388       & 0.313            &  0.356             \\   
16       &  72.14        & 0.412       & 0.450       & 0.375            &  0.416             \\   
5        &  72.21        & 0.471       & 0.472       & 0.438            &  0.437             \\   
13       &  72.33        & 0.529       & 0.509       & 0.500            &  0.475             \\   
17       &  72.39        & 0.588       & 0.528       & 0.563            &  0.494             \\   
2        &  72.47        & 0.647       & 0.553       & 0.625            &  0.519             \\   
1        &  72.53        & 0.706       & 0.572       & 0.688            &  0.538             \\   
7        &  72.64        & 0.765       & 0.606       & 0.750            &  0.572             \\   
6        &  72.65        & 0.824       & 0.609       & 0.813            &  0.575             \\   
HV004    &  73.23        & 0.882       & 0.768       & 0.875            &  0.740             \\   
9        &  73.50        & 0.941       & 0.828       & 0.938            &  0.805             \\   
HV055    &  73.62        & 1.000       & 0.851       & 1.000            &  0.830             \\   
\hline    
   
\end{tabular}   
\end{scriptsize}   
\end{table}

\section{Method}

Given the small sample size, the formal error on the velocity dispersion is not
sufficient to discriminate between various models, so we use the KS-test to 
redo the analysis of Jordi et al.~(2009).
The KS-test compares two cumulative distribution functions (cdfs), then 
the maximum difference D between these two cdfs yields a P-value. 

Here we compare the cdf of the data (separately for the samples with 
and without Star 15) to the cdf of a Gaussian with dispersion equal to
1.27 km s$^{-1}$ (the MOND prediction with M/L=2, Baumgardt et al. 2005, 
Jordi et al. 2009). We note
that there is no evidence, independent from dynamics itself, about the centre of the Gaussian
therefore we chose the centres that minimise D (in other words, the
centres that maximise the P-value).

The MOND prediction for the velocity dispersion was derived by Baumgardt et al. (2005). 
First, they calculated the internal and external field gravitational accelerations 
($a_{\rm int}$ and $a_{\rm ext}$, respectively) of a number
of globular clusters, including Pal~14. Then, based on results 
obtained by Milgrom (1986), they found that the MOND prediction is 
simply given by the Newtonian one multiplied by $\sqrt{a_{0}/a_{\rm ext}}$,
because Pal~14 is in deep MOND regime and $a_{\rm ext}$ is larger than  $a_{\rm int}$.
Baumgardt et al. (2005) assume M/L=2 from the observed M/L ratios of globular clusters:
Mandushev et al. (1991) find M/L=1.21 from 32 clusters, whereas Pryor \& Meylan (1993) find M/L=2.3
from 56 clusters. Baumgardt et al. (2005) state that the latter value is more plausible because
the modelling used in Pryor \& Meylan (1993) takes mass segregation into account.  
To get an estimate of the uncertainty on the M/L, the data compilation by Dabringhausen et al. 
(2008) suggest that M/L ratios of globular clusters have a spread of roughly 0.5 dex.

\section{Results}

\begin{figure}
   \centering
   \includegraphics[width=8.2cm]{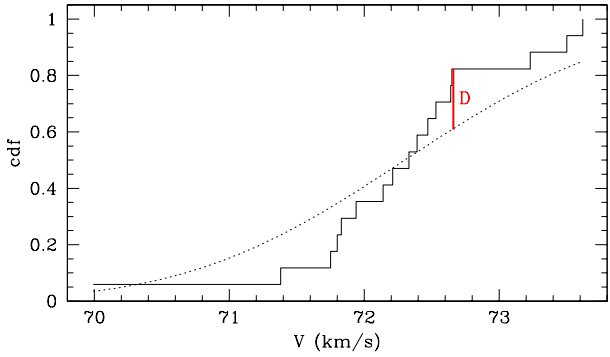}
   \caption{Cumulative distribution function (cdf) vs. radial velocity for
the sample with Star 15. The solid step-like line represents the data,
whereas the dotted line is the cdf of a Gaussian with standard deviation of
1.27 km s$^{-1}$ (the MOND prediction) and centre at 72.30 km s$^{-1}$.
The red segment shows $D$, the maximum difference between the two cdfs.  
}
   \label{ks}
    \end{figure}

\begin{figure}
   \centering
   \includegraphics[width=8.2cm]{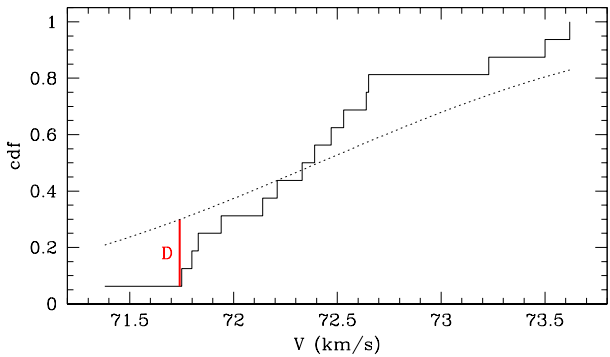}
   \caption{Cumulative distribution function (cdf) vs. radial velocity for
the sample without Star 15. The lines are the same as in Fig. \ref{ks},
but the Gaussian's centre is at 72.41 km s$^{-1}$.
}
   \label{ks_no15}
    \end{figure}

In Table \ref{tab-cdf} we list the observed and predicted cdfs 
(Gaussians with a standard deviation of 1.27 km s$^{-1}$), separately
for the sample with and without Star 15. The two predicted cdfs are
different because they have different centres (72.30 km s$^{-1}$ for the
sample with Star 15 and 72.41 km s$^{-1}$ for the
sample without Star 15). We also show them in Figs. \ref{ks} and \ref{ks_no15}.
In both cases, there are two regions where the difference goes close to the 
maximum: around 71.75 km s$^{-1}$ and around 72.65 km s$^{-1}$, respectively. 
And in both
samples, the difference between the two observed and predicted cdf are
very similar at these two velocities. With Star 15, the maximum difference
is 0.215, and without Star 15 the maximum difference is 0.239.
These values of $D$ correspond to P-values of 0.360 and 0.273, respectively.
This means that using the KS-test, the data presented in Jordi et al. (2009) 
can exclude the MOND with M/L=2 hypothesis only with 64\% and 73\% confidence, depending on the inclusion of Star 15. These confidence levels are clearly not sufficient to exclude MOND.
We note that if the M/L is not 2, but rather 1.1 (a possibility considered by Jordi et al. 2009), then MOND is perfectly consistent with a velocity dispersion of 0.85 km s$^{-1}$.

\begin{figure}
   \centering
   \includegraphics[width=8cm]{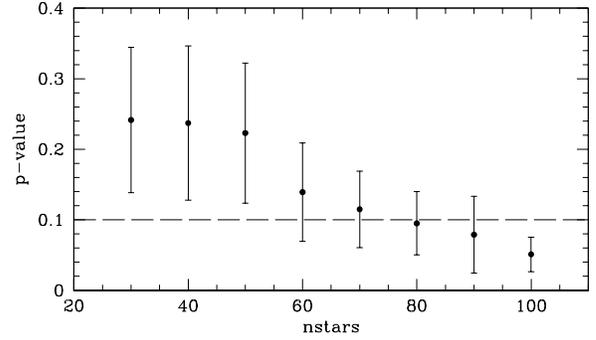}
   \caption{Mean P-value vs. number of stars in the sample for a number of Monte-Carlo realizations
(10 per sample size) of distributions of stars following a Gaussian with a 
standard deviation of 0.85 km s$^{-1}$ (equivalent to the sample with Star 15). }
   \label{nstars}
    \end{figure}

\begin{figure}
   \centering
   \includegraphics[width=8cm]{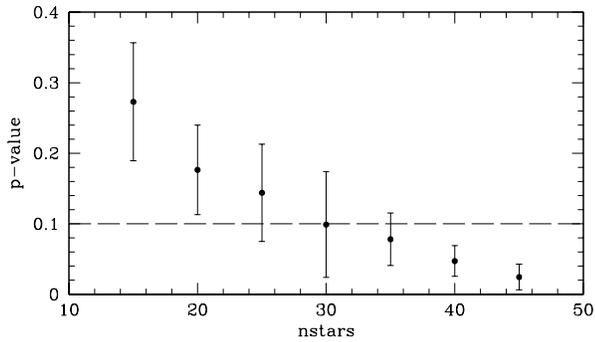}
   \caption{Same as in Fig. \ref{nstars}, but for a Gaussian with a standard
deviation of 0.63 km s$^{-1}$ (equivalent to the sample without Star 15).}
   \label{nstars_no15}
    \end{figure}

Now we can ask ourselves: if one wanted a P-value of 0.1
(i.e. an exclusion confidence of 90~\%), how many stars would
one need (with the hypothesis that they follow a Gaussian with standard
deviation equal to the measured one)? We measure a dispersion of 0.85 km s$^{-1}$
with Star 15 and 0.63 km s$^{-1}$ without Star 15. Hence, we create a series of 
Monte-Carlo realizations, using the software package R (Ihaka \& Gentleman 1996), of increasing sample size.
We create 10 realizations per sample size.
Then, we compare these mock data sets to the MOND prediction using the same method
as above, take the mean P-value 
for each sample size, and look for the minimum number of stars such that 
P $\le$ 0.1 is obtained. Our results (shown in Figs. \ref{nstars}
and \ref{nstars_no15}) are that for the mock data sets
equivalent to the sample with Star 15 (i.e., the mock data sets that follow a Gaussian
with standard deviation = 0.85 km s$^{-1}$) a minimum of about 80 stars
(or alternatively 80/17 $\sim$ 5 clusters) would have been needed
to exclude the MOND hypothesis with a confidence level of 90\%, whereas without star
15 this minimum number of stars decreases to about 30.

\section{Discussion}

In this research note, we showed that current observational data on Pal~14 are not significantly discrepant with the theoretical prediction of MOND. Let us however also note that this theoretical prediction is not really unique, and depends on many factors, such as stellar M/L, rotation and anisotropy. For instance, in Angus (2008, thesis), the velocity dispersion of Pal~14 was discussed before any velocity dispersion was published: the line of sight velocity dispersion for various M/Ls and velocity anisotropies were computed (radial, isotropic and centrally isotropic with increasing tangentially biased orbits). It was found that lower velocity dispersions (consistent with 0.85~km$\, {\rm s}^{-1}$) were attainable with very radial orbits, and in that case the line of sight velocity dispersion profile would moreover not be flat in the outskirts. Rotation could also play a role although the round appearance of Pal~14 justifies the no-rotation assumption. 
Of course, the most important uncertainty comes from the M/L ratio itself.

Then, another possible oversight may be the mass segregation of stars within the globular cluster: if significant energy partitioning has occured due to the short MOND relaxation time (Ciotti \& Binney 2004), this would mean that the low-mass, unobserved, stars could have a larger spread in configuration space and a higher velocity dispersion. Also, as already noted in Jordi et al.~(2009), the MOND theoretical prediction was based on assuming a purely circular orbit for Pal~14 around the Milky Way, while if it is on an eccentric orbit, it could (i) have lost many low mass stars at perigalacticon leading to a decrease of the theoretical stellar M/L, and (ii) be ``frozen" in the Newtonian regime due to the period of recovery while transiting from the large external field endured at perigalacticon. 
Finally, we note that the fact that MOND cannot be excluded also 
implies, in the Newtonian plus dark matter cosmological framework, that the current data of Pal 14 cannot exclude a certain amount of unseen mass in the cluster. With the assumption of a stellar M/L of 2,
a velocity dispersion of 1.27 km s$^{-1}$ would imply a total-to-luminous mass ratio of 2 to 4.

\section{Conclusion}

Even assuming an isotropic MOND model with no rotation for a globular cluster on a circular orbit with M/L=2, we showed that, based on a KS test, which is the relevant statistical test for small samples, the currently available data are insufficient to discriminate between Newtonian gravity and MOND in Palomar~14, contrary to the claim of Jordi et al.~(2009). While the objects proposed by Baumgardt et al.~(2005) provide one of the 
best discriminating test between MOND and cold dark matter plus Newtonian 
dynamics, more observations would be needed to exclude the aforementioned MOND model if the observed velocity dispersion is representative of the true one: about 80 stars in Pal~14, or about 5 similarly problematic globular clusters for this model. In this respect, even though it might still not be conclusive, velocity data of 21 stars in Pal~3 and 24 stars in Pal~4 will be of prime interest for testing fundamental physics in our Galactic backyard.

\begin{acknowledgements}
GG is a postdoctoral researcher of the FWO-Vlaanderen (Belgium). BF acknowledges support from the Hulmboldt foundation and the CNRS. GWA is supported by the Univsersity of Torino and Regione Piemonte, and by the INFN grant PD51.
\end{acknowledgements}


\begin{thebibliography}{}
\bibitem[]{AFD} Angus, G., Famaey, B. \& Diaferio, A.\ 2009, arXiv:0906.3322
\bibitem[Baumgardt et al.(2005)]{2005MNRAS.359L...1B} Baumgardt, H., Grebel, E.~K., \& Kroupa, P.\ 2005, MNRAS, 359, L1 
\bibitem[Baumgardt et al.(2009)]{2009MNRAS.396.2051B} Baumgardt, H., et al.\ 2009, MNRAS, 396, 2051 
\bibitem[]{CB} Ciotti, L., \& Binney, J.\ 2004, MNRAS, 351, 285
\bibitem[Dabringhausen et al.(2008)]{2008MNRAS.386..864D} Dabringhausen, J., Hilker, M., \& Kroupa, P.\ 2008, 
MNRAS, 386, 864 
\bibitem[]{don} Donato, F., et al.\ 2009, MNRAS, 397, 1169
\bibitem[]{FJ} Faber, S., \& Jackson, R.\ 1976, ApJ, 204, 668
\bibitem[Gentile et al.(2007)]{2007A&A...472L..25G} Gentile, G., et al.\ 2007, A\&A, 472, L25 
\bibitem[]{urc} Gentile, G.\ 2008, ApJ, 684, 1018
\bibitem[]{nature} Gentile, G., Famaey, B., Zhao, H.~S., \& Salucci, P.\ 2009, Nature, 461, 627
\bibitem[Harris \& van den Bergh(1984)]{1984AJ.....89.1816H} Harris, W.~E., \& van den Bergh, S.\ 1984, AJ, 89, 1816 
\bibitem[Hilker(2006)]{2006A&A...448..171H} Hilker, M.\ 2006, A\&A, 448, 171 
\bibitem[Hinshaw et al.(2009)]{2009ApJS..180..225H} Hinshaw, G., et al.\ 2009, ApJS, 180, 225 
\bibitem[Holland \& Harris(1992)]{1992AJ....103..131H} Holland, S., \& Harris, W.~E.\ 1992, AJ, 103, 131 
\bibitem[1996]{r-project} Ihaka, R., \& Gentleman, R., 1996, Journal of Computational and Graphical Statistics, 5(3),299,  URL: http://www.r-project.org/.
\bibitem[Jordi et al.(2009)]{2009AJ....137.4586J} Jordi, K., et al.\ 2009, AJ, 137, 4586 
\bibitem[Mandushev et al.(1991)]{1991A&A...252...94M} Mandushev, G., Staneva, A., \& Spasova, N.\ 1991, A\&A, 252, 94 
\bibitem[]{McG} McGaugh, S.~S., et al.\ 2000, ApJ, L99
\bibitem[]{MDA} McGaugh, S.S.\ 2004, ApJ, 609, 652
\bibitem[Milgrom(1983)]{1983ApJ...270..365M} Milgrom, M.\ 1983, ApJ, 270, 365 
\bibitem[Milgrom(1986)]{1986ApJ...302..617M} Milgrom, M.\ 1986, ApJ, 302, 61
\bibitem[]{Ols} Olszewski, E.~W., Pryor, C., \& Shommer, R.~B., ASP Conf. Series, 48, 99
\bibitem[Pryor \& Meylan(1993)]{1993ASPC...50..357P} Pryor, C., \& Meylan, G.\ 1993, Structure and Dynamics of Globular Clusters, 50, 357
\bibitem[Sanders(1999)]{1999ApJ...512L..23S} Sanders, R.~H.\ 1999, ApJ, 512, L23 
\bibitem[]{SN} Sollima, A., \& Nipoti, C.\ 2009, arXiv:0909.1656
\bibitem[Soong(1994)]{soong} Soong, T. T., 2004, Fundamentals of Probability and Statistics for Engineers. Publisher: John Wiley \& Sons
\bibitem[]{TF} Tully, R.~B. \& Fisher, J.~R. 1977, A\&A, 54, 661


\end{thebibliography}
\end{document}